# Agile Meets Quantum: A Novel Genetic Algorithm Model for Predicting the Success of Quantum Software Development Project

Arif Ali Khan[1], Muhammad Azeem Akbar[2], Valtteri Lahtinen[3], Marko Paavola[4], Mahmood Niazi[5,6], Mohammed Naif Alatawi[7], Shoayee Dlaim Alotaibi[8]

[1]M3S Empirical Software Engineering Research Unit, University of Oulu, 90014 Oulu, Finland
[2]Software Engineering Department, Lappeenranta-Lahti University of Technology, 53851 Lappeenranta, Finland
[3]Quanscient Oy, Tampere, Finland
[4]VTT Technical Research Ctr of Finland Ltd, Oulu, Finland
[5]Department of Information and Computer Science, King Fahd University of Petroleum and Minerals, Saudi Arabia
[6]Interdisciplinary Research Centre for Intelligent Secure Systems, King Fahd University of Petroleum and Minerals, Saudi Arabia
[7]Information Technology Department, Faculty of Computers and Information Technology University of Tabuk, Saudi Arabia
[8]Department of Artificial Intelligence and Data Science, College of Computer Science and Engineering, University of Hail , Hail, Saudi Arabia
arif.khan@oulu.fi, azeem.akbar@lut.fi, valtteri.lahtinen@quanscient.com, marko.paavola@vtt.fi, mkniazi@kfupm.edu.sa, alatawimn@ut.edu.sa, s.alotaibi@uoh.edu.sa

**Abstract**

**Context:** Quantum software systems represent a new realm in software engineering, utilizing quantum bits (Qubits) and quantum gates (Qgates) to solve the complex problems more efficiently than classical counterparts. Agile software development approaches are considered to address many inherent challenges in quantum software development, but their effective integration remains unexplored.

**Objective:** This study investigates key causes of challenges that could hinders the adoption of traditional agile approaches in quantum software projects and develop an Agile-Quantum Software Project Success Prediction Model (AQSSPM).

**Methodology:** Firstly, we identified 19 causes of challenging factors discussed in our previous study, which are potentially impacting agile-quantum project success. Secondly, a survey was conducted to collect expert opinions on these causes and applied Genetic Algorithm (GA) with Naive Bayes Classifier (NBC) and Logistic Regression (LR) to develop the AQSSPM.

**Results:** Utilizing GA with NBC, project success probability improved from 53.17% to 99.68%, with cost reductions from 0.463% to 0.403%. Similarly, GA with LR increased success rates from 55.52% to 98.99%, and costs decreased from 0.496% to 0.409% after 100 iterations. Both methods result showed a strong positive correlation (rs=0.955) in causes ranking, with no significant difference between them (t=1.195, p=0.240>0.05).

**Conclusion:** The AQSSPM highlights critical focus areas for efficiently and successfully implementing agile-quantum projects considering the cost factor of a particular project.

## 1. Introduction

Quantum Computing (QC) has emerged as a significant area of interest for industrial practitioners and academic researchers worldwide. Its potential to bring transformative changes across various industries is undeniable [1]. This potential is evident in the investments and efforts of technology leaders such as IBM [2], Google [3], and Microsoft [4], who are actively working to harness QC for solving complex computational challenges. However, as QC machines operate on the principles of quantum mechanics, the development of software applications for them presents unique challenges [5]. Literature suggests that the development process for QC applications closely mirrors traditional software development [1], underscoring the importance of a well-structured engineering process to address the complexities of QC [1]. Therefore, refined Software Engineering (SE) methodologies are necessary to fully realize the benefits of QC [1, 5, 6].

A key component in operationalizing QC systems is quantum software [7]. This software, requiring comprehensive stack support and innovative tools and techniques, demands specific processes and methods tailored to create software systems based on quantum mechanics principles [7]. However, the field of Quantum Software Engineering (QSE) is still evolving. Currently, it primarily relies on hybrid (quantum and classical) processes, tools, and methods, necessitating careful coordination between classical and quantum systems. In this regard, we previously conducted an empirical study introducing the idea of utilizing traditional (classical) iterative and agile methodologies for developing quantum software [8, 13]. These agile methods, which promote team collaboration, quick iterations, and continuous delivery, could significantly benefit QSE activities [8, 9, 10]. Nevertheless, more evidence is needed to directly connect agile methods to quantum software development.

Drawing from our previous research, where we interviewed 26 software professionals from 10 different countries [8], we found that many view existing traditional agile methods as well-suited for quantum software development. They also identified potential challenges perceived to hinder scaling traditional agile methods in the quantum domain. These challenges fall into four main categories: 1) *Knowledge and awareness*, offering practitioners a deeper understanding of potential changes that might impede the successful adoption of agile methodologies in quantum software development; 2) *Sustainable scaling*, involving challenges related to scaling software sustainably and ethically; 3) *Quantum-aware tools and technologies*, encompassing challenges related to technological infrastructure that could support customizing, automating, and tailoring agile practices for developing quantum software; and 4) *Standards and specifications*, providing insights into challenges related to the lack of standards, frameworks, and common rules for adopting existing agile approaches in the quantum software domain.

With these findings, this paper aims to extend the previous study [8] by exploring more in-depth the causes behind these challenges and providing effective best practices to mitigate them. Based on the study results, we will introduce a model to predict the costs and efforts required for scaling agile methods in the quantum software development domain. This model utilizes a nature-inspired optimization algorithm, the Genetic Algorithm (GA) [11], to evaluate the likelihood of success in terms of cost. Within the GA, we designed a measure called the fitness function, aiming for the highest success rate of agile quantum software projects while considering costs. We tested this optimization model using two prediction models: the Naive Bayes Classifier (NBC) and Logistic Regression (LR). Through this comprehensive approach, we aim to provide a roadmap for both researchers and practitioners in the quantum software development field, emphasizing the integration of agile practices.

## 2. Background

In the rapidly evolving world of technological advancements, QC has emerged as a pivotal innovation [1]. Based on the principles of quantum mechanics, such as quantum superposition and entanglement, QC platforms offer a new computational paradigm. Unlike traditional binary digits, which strictly adhere to a 0 or 1 state, QC operates based on Qubits that can simultaneously exist in a superposition of both states, represented as |0 and |1 [1, 12]. This shift from classical bits to Qubits signifies a profound transformation in the landscape of computing, pushing the boundaries of what is achievable with computational processes [1].

To navigate this quantum realm effectively, software developers are required to be familiar with quantum-age algorithms and programming languages like Microsoft Q# [4], IBM Qiskit [2], and Google Cirq [3]. However, there is a lack of a holistic engineering approach integral for quantum software development, e.g., quantum software requirements engineering, quantum software design, maintenance, evolution, and information simulation [13, 14]. This gap underscores the importance of Quantum Software Engineering (QSE), as highlighted in a national agenda for software engineering research and development presented by SEI-Carnegie Mellon University [15]: "*the development of Quantum Computing Software Systems is considered a pivotal future focus area. The objectives of this research area are initially focused on facilitating the development of quantum software for existing quantum computers and then enable increasing abstraction as larger, fully fault-tolerant quantum computing systems become available. A key challenge is to eventually, fully integrate these types of systems into a unified classical and quantum software development life cycle.*" As the most novel field, QSE seeks to leverage the best practices from traditional software engineering, such as robust engineering processes, architectures, and patterns, facilitating the development of quantum-intensive software solutions. There is no existing quantum-specific process model for QSE available, except for a few research studies which focused on defining the fundamental engineering activities for a structured QSE approach [13, 14, 16].

In the seek for structure and a robust QSE process, the agile approach from traditional software engineering presents itself as a promising contender. Tracking the evolution of traditional software programming, which began with a strong focus on hardware in the 1950s and later shifted to today's agile practices [17, 18], opens the door for the novel QSE field to adopt traditional agile methods over time. However, given the unique challenges of quantum software development, such as qubit entanglement and superposition, it is increasingly compelling to examine the suitability of traditional agile approaches in line with the characteristics of quantum software development [19-21]. For example, composing, designing, testing, and maintaining quantum programs would be challenging due to switching to an entirely different mindset, examining the superposition states of quantum variables, and interpreting multi-dimensional quantum states.

In traditional software development, the above-mentioned activities are addressed by using agile and iterative approaches. Agile methods, in contrast to traditional models such as Waterfall, are considered best suited to propose practices and principles such as active user involvement, iterative development, frequent releases, continuous integration, and deployment [9]. However, the QSE field is still in its infancy, and the process of creating quantum software can be made simpler by embracing agile methods, as recommended by Piattini et al. [18] and Khan et al. [8].

Following this, we previously conducted an empirical study to evaluate the suitability of the traditional agile approaches and identify the challenges that could be potential barriers for agile-based quantum software development [8]. In our prior study, we reached the conclusion that agile approaches are the preferred choice for developing quantum software. The study also brought to light a range of challenges that could be potential barriers to adopting agile practices. Based on our earlier work, this research aims to investigate deeper into the causes of these challenges and best practices in-depth to formulate a predictive model that will estimate the chances of success when traditional agile techniques are used in quantum software projects.

No previous research has been conducted on developing a prediction model to aid decision-making about the use of agile practices in the quantum software domain. We have set out to address this gap by proposing a model that we name AQSSPM. This proposed model will serve as a tool for organizations that are considering the adoption of agile methods for quantum software development, highlighting the most significant key areas that can predict the success of their endeavors in this new and complex field.

For the creation of this model, we have incorporated not just the findings from our previous studies [8], but also an expanded investigation into the causes behind these challenges and the most effective practices to address them. An industrial survey was conducted to gather practitioners' views on the challenges, causes, and best practices in quantum software development as identified in our previous research. The feedback obtained from this survey has been crucial in applying genetic algorithms to develop the proposed model - AQSSPM.

**3. Research Methodology**
In this study we used the survey questionnaire approach to collect the data from industrial practitioners to evaluate the causes of the identified challenges. Based on the collected data, we trained the GA with NBC and LR to estimate the success probability of agile-quantum software projects. The details are provided in the following sections:

**3.1 Designing the Questionnaire**
To collect the training data, we initiated data collection process through unstructured interviews with 13 experts in the domains of quantum computing and agile software development. These conversations, facilitated via digital meeting platforms like Google Meet and Zoom, spanned an average of 25 minutes each. Building upon the insights derived from these interactions, we constructed a detailed, closed-ended survey. The questionnaire survey allowed us to pinpoint 19 causes contributing to the challenges faced in agile-quantum software development. We used the questionnaire survey methodology to collect the data from a broad and varied group, working in agile-quantum paradigm.

The survey was divided into two sections. The first section describes the demographic details of participants, while the subsequent part contains closed-ended questions centered around the core variables (causes). To assess the feedback, we used a 9-point likert scale, range from extremely low (EL) to extremely high (EH). Using surveys for data collection, especially when observational techniques fall short, is a best suited approach, as evidenced in different other studies [22, 23].

**Pilot Assessment of the Questionnaire**
To enhance the accuracy and consistency of our survey insturment, we conducted a pilot assessment of the questionnaire. As highlighted in existing studies, such pre-tests are important to refine the survey structure and ensure relevant feedback [24, 25]. For this early assessment, we involved ten specialists, of which six participated in our prior informal discussions and four were new contributors. These specialists were associated with research institutions, including the University of Oulu in Finland, Norwegian University of Science and Technology in Norway, and Aligarh Muslim University in India.
Based on the feedback received, we updated the questionnaire structure by classifying it into three sections: demographic details, specific questions about identified causes of agile-quantum project, and cost-related data for tackling the identified causes. We also adjusted the wording of the survey items for clarity. One expert suggested presenting the questions in a tabular format, and other asked to improve the understandability of each question. The refined questionnaire can be accessed at: *https://tinyurl.com/43r7uxx6*.

**3.2 Process of predictive model development**
The presented study seeks to understand the impact of pinpointed causes by identifying the scale at which the project has the highest likelihood of success. This insight will guide software professionals to prioritize critical elements in the agile-quantum workflow, ensuring its optimal operation. To achieve this, we employed Genetic Algorithms (GA) [11] aiming to optimize the chances of project success while restricting the related costs. GA works by generating factor states, determining success probabilities with predictive models, and simultaneously evaluating the balance between success likelihood and cost, thereby pinpointing feasible factor scales. The efficiency of GA hinges on three main components: (1) Predictive models based on Naive-Bayes and logistic regression, which inform GA of the success probability aligned with specific risk factor values; (2) The expense linked to each risk factor's scale value; (3) A function grounded in success probability and cost, guiding GA about the merit of a certain set of feature values. In this context, once the predictive models are adequately trained, they provide GA with the necessary probability, with our primary focus on the Naive Bayes algorithms. The cost metrics for different process areas, derived from expert opinions, are detailed in Table 1. Lastly, an efficiency function is crafted to meet the third component's needs, taking into account both success probability and cost.

Table 1: Causes at different scales, along with associated costs

| Scale | C1 | C2 | C3 | C4 | C5 | C6 | C7 | C8 | C9 | C10 | C11 | C12 | C13 | C14 | C15 | C16 | C17 | C18 | C19 |
|---|---|---|---|---|---|---|---|---|---|---|---|---|---|---|---|---|---|---|---|
| EL | 2 | 2 | 2 | 2 | 1 | 1 | 1 | 1 | 3 | 1 | 1 | 2 | 2 | 4 | 3 | 1 | 2 | 1 | 1 |
| VL | 3 | 2 | 2 | 2 | 2 | 2 | 1 | 1 | 3 | 2 | 2 | 3 | 3 | 3 | 3 | 1 | 2 | 4 | 2 |
| L | 3 | 2 | 3 | 3 | 2 | 2 | 2 | 2 | 4 | 2 | 3 | 3 | 3 | 2 | 4 | 2 | 4 | 3 | 2 |
| SL | 4 | 4 | 4 | 4 | 2 | 3 | 3 | 3 | 4 | 2 | 4 | 4 | 4 | 4 | 4 | 5 | 4 | 5 | 2 |
| Neutral | 5 | 5 | 4 | 4 | 4 | 4 | 5 | 4 | 5 | 5 | 5 | 5 | 5 | 4 | 5 | 4 | 5 | 5 | 4 |
| SH | 6 | 6 | 5 | 6 | 6 | 6 | 6 | 4 | 5 | 6 | 6 | 6 | 6 | 6 | 5 | 3 | 6 | 6 | 6 |
| MH | 6 | 6 | 6 | 6 | 6 | 6 | 6 | 5 | 6 | 6 | 7 | 7 | 7 | 7 | 6 | 5 | 7 | 7 | 8 |
| VH | 7 | 7 | 7 | 7 | 7 | 7 | 6 | 6 | 7 | 6 | 7 | 8 | 8 | 7 | 7 | 6 | 7 | 8 | 7 |
| EH | 8 | 8 | 8 | 8 | 7 | 7 | 7 | 8 | 8 | 7 | 8 | 9 | 9 | 8 | 8 | 8 | 8 | 9 | 7 |

**Predictive models**
In our research, we employed predictive models grounded in Naive Bayes and Logistic Regression methodologies. These models produce the likelihood of a class variable reaching a particular outcome, representing success or failure in our context.

**Naive Bayes Classifier (NBC)**
The NBC model estimates the probability of a specific result for a class variable, like success or failure. Bayesian Networks offer an array of classifiers rooted in probability [26]. Notably, the Naive Bayes Classifier (NBC) is distinguished by its straightforwardness and efficacy [27].

The NBC presumes that, given the class, the predictors or attributes function independently. In this framework, every independent variable possesses a sole parent: the class or outcome variable. Because of its robust mathematical underpinning, the NBC algorithm is renowned for its rapidity, simplicity in execution, and adaptability to manage datasets with vast dimensions. Such effectiveness stems mainly from NBC approach of determining the probability for each attribute autonomously [27, 28].

Equation 1 depicts the computation of the maximum posterior probability for the target variable T in relation to the attributes F of the observation, grounded in Bayes' theorem:

$$Prob(T|F) = \frac{Prob(T)*Prob(F|T)}{Prob(F)} \quad (1)$$

NBC considers that all the elements of F = {f$_1$, f$_2$,…,f$_n$}, given T, are conditionally independent, and therefore, the probability described in Equation 1 can be computed according to Equation 2.

$$Prob(T|F) = \frac{Prob(T) \prod_{i=1}^{n} Prob(f_i|T)}{Prob(F)} \quad (2)$$

Equation 2 can be rewritten as Equation 3 in its extended form:

$$Prob(T|f_1, f_2, \ldots f_n) = \frac{Prob(T)*Prob(f_1|T)*Prob(f_2|T)\ldots *Prob(f_n|T)}{Prob(F)} \quad (3)$$

For categorization endeavors, Equation 3 often suffices to pinpoint the likeliest state of the target variable based on a defined set of factors. Yet, in our study, we harness the Naive Bayes methodology, utilizing input values from diverse factors to forecast the project's success probability. After the model's training phase, it's equipped to gauge the potential for a positive result.

**Logistic regression (LR)**

In our study, we additionally utilize logistic regression (LR) as a forecasting tool, mainly to determine the likelihood of a binary outcome [28]. This is perceived as an evolution of regression methods tailored for predicting continuous outcome variables. A drawback with conventional regression techniques is that they can produce predicted values for the outcome variable that fall outside the (0,1) range, with 0 representing a negative outcome (Failure, False, or No) and 1 a positive outcome (Success, True, or Yes) [29]. To address this challenge, logistic regression employs a logistic function, as presented in Equation (4):

$$S(x) = \frac{1}{1+ e^{-x}} \quad (4)$$

Initially, a function is defined based on independent variables as follows:

$$func(F) = \beta_0 + \beta_1 f_1 + \ldots + \beta_n f_n \quad (5)$$

In Equation (5), $\beta_i$ denotes the significance of attribute f$_i$. The core goal of the Logistic Regression (LR) algorithm is to determine the best-fit values for each $\beta_i$. Using Equation (6), LR produces probabilistic estimates, categorizing the binary target variable T into either 1 or 0.

$$Prob(T = 1) = \frac{1}{1+ e^{-func(F)}} \quad (6)$$

Equation (7) can be used to calculate the probability ($T == 0$):

$$Prob(T=0) = 1 - Prob(T=1) \quad (7)$$

The LR algorithm executes a series of procedures to refine the $\beta_i$ values according to Equations (5) and (6) until a state of relative balance is achieved, where the $\beta_i$ values remain largely consistent. At this stage, the LR algorithm integrates the attributes to optimize the likelihood of discerning the state of T in terms of probability. In our research context, we employed both NBC and LR models. These models intake the input values from diverse factors to forecast the likelihood of project success. Once honed, these models are adept at gauging the potential for a favorable result.

## Optimization problem

This segment exmined the mathematical framing of the optimization challenge, paving the way for the implementation of the Genetic Algorithm (GA). The predictive models are primed to compute probability figures using the supplied data. Definitions for both probability and cost will be established, and from these metrics, an effectiveness function will be subsequently formulated.

## Probability of success

Based on a given set of attributes, a project's probability of success can be expressed as follows:

$$\text{Prob}(S) = p \qquad (8)$$

Prob (S) takes in a prospective solution S and evaluates its success probability p, ranging between 0 and 1. The set S can be articulated as follows:

$$S = \{s_1, s_2, \ldots, s_i, \ldots, s_n\} \qquad (9)$$

In Equation 9, the variable $s_1$ represents the scale of the $i^{th}$ factor, with n indicating the aggregate count of factors being examined. Here, $s_i$ can take on values from 1 through 9, and n matches the count of these factors. Equation 10 showcases a particular example of the solution set S, denoted as S (n=19).

$$S' = \{6,5,4,2,3,7,1,3,8,2,6,1,9,6, 8\}1 \qquad (10)$$

In this case, S' represents a solution set in which the first factor has a scale value of 6, the second factor holds a value of 5, and so on. Introducing S' as an input into the model produces the associated probability, given that the predictive models have been previously trained.

## Cost Calculation

An important section of defining the problem is identifying the costs tied to each factor's scale. These costs have been determined manually by domain experts. The objective is to amplify the likelihood of project success while constraining the overall expenses. The variable $c_{ij}$ represents the cost associated with the $j^{th}$ scale value of variable i. The cumulative cost for the solution S can be determined through Equation 11.

$$C(S) = \sum_{i=1}^{n} c_{ij} \qquad (11)$$

Table 1 outlines the expenses linked to different scales of factors. Whenever a fresh instance of S is produced, the overall project cost is ascertained based on Table 1 in conjunction with Equation (11).

## Efficacy

A project's effectiveness is determined by evaluating its likelihood of success alongside its associated cost, see Equation 12. This challenge can be perceived as a dual-objective optimization task, striving to enhance the probability of project success while simultaneously curbing associated expenditures. To merge these two objectives into one optimization dilemma, an effectiveness function was crafted, as detailed below:

$$E = \text{Prob} - C \qquad (12)$$

One practical approach to gauge the effectiveness of a particular instance S is by determining "the disparity between the success probability and its cost." This straightforward method integrates both parameters into a singular function. Yet, as illustrated in Equation 8, the cost C often dominates, given that the probability Prob consistently lies between [0,1], whereas C might attain a peak value of

max(C) when all factors are scaled at 9. To address this discrepancy, the normalized cost, as presented in Equation 13, can be utilized.

$$norm(C) = \frac{C - \min(C)}{\max(C) - \min(C)} \quad (13)$$

In Equation 13, C denotes the cost awaiting normalization, while min(C) and max(C) represent the minimum and maximum costs for a project, respectively. In the context of our problem, there are 19 causes (meaning, n=19). As such, max (C) is 126, and min (C) stands at 19 (assuming all factors are scaled at 1). Utilizing Equation 13, we confine the cost within the range (0,1), guaranteeing it doesn't solely influence Equation 12. The consequent effectiveness of S is then determined using Equation 14.

$$E(S) = Prob(S) - norm(C(S)) \quad (14)$$

It's pertinent to mention that various approaches exist to integrate the cost into the problem formulation. One approach is to treat both cost and success probability as separate objectives, which would morph the present issue into a multi-objective optimization challenge. Another perspective would be to perceive the cost as a constraint rather than a component of the fitness function. In this scenario, the aim would be to identify a solution with the utmost success probability, given that the cost remains below a predetermined maximum allowable expense, represented as $C_{max}$. However, for its straightforwardness and lucidity, this research opted to use Equation 14 as the guiding objective function.

**Mathematical modeling of the optimization problem**
After detailing the essential components of the problem in the previous sections, we now present the optimization challenge and its corresponding mathematical representation, which is to be maximized.

$$\text{Maximize } E(S) = Prob(S) - norm(C(S)) \quad (15)$$
$$\text{where } S = \{s_1, s_2, \ldots, s_i, \ldots, s_n\}$$
$$\text{Given: } s_{min} \leq s_i \leq s_{max}$$

In the given equation, $S_{max}$ and $S_{min}$ denote the maximum and minimum scale values, respectively. From Equation 15, it is evident that our goal is to identify an instance S that yields the optimal effectiveness value, which is influenced by a high probability of success paired with a minimized normalized cost.

**Optimization Problem, Genetic Algorithm, and Its Significance**
While traditional solutions like exhaustive search can address optimization problems, they become impractical when dealing with extensive search spaces [30]. In our current context, with fourteen attributes, each potentially ranging from 1 to 9, we encounter over 22.8 trillion potential solutions ($9^{19}$ > 22.8 trillion). Hence, meta-heuristic methods, like GA, are favored as they can deliver near-optimal results within acceptable timeframes. GA stands out as a prevalent meta-heuristic strategy for tackling optimization challenges, with numerous scholars employing it across various domains, especially in combinatorial optimization tasks. While other meta-heuristics exist, the No-Free-Lunch Theorem [29] suggests that no singular metaheuristic outshines the rest, and overall, they yield comparable results. Numerous studies have showcased GA's application across diverse sectors for combinatorial optimization issues, consistently yielding encouraging outcomes [31-33, 34]. Therefore, this research employs GA.

**Genetic Algorithm**

The Genetic Algorithm (GA) is an evolutionary technique rooted in Charles Darwin's theory of natural selection and was formulated by John Holland in the 1970s [35]. It leverages the principle of natural selection to choose potential parent solutions from a population, with the aim of generating superior offspring in subsequent generations [35]. GA refines the set of solutions with each iteration, progressively nearing the optimal solution. It proves especially potent when dealing with objective functions that are stochastic, non-linear, non-differentiable, or discontinuous. GA is anchored on three core operations: selection, crossover, and mutation. These operations guide the generation of the optimal solution with each successive iteration [11]. The foundational steps of the standard GA technique are showcased in Algorithm 1.

---

**Algorithm 1:** Standard Genetic Algorithm
**Step 1:** Randomly generate the initial population of chromosomes, say P.
**Step 2:** Calculate fitness function f (x) for each chromosome x in P.
**Step 3:** Generate child population C by applying selection, crossover, and mutation operators on P.
**Step 4:** P ← C.
**Step 5:** Repeat Steps (2-4) until convergence.

---

Fine-tuning the elements of GA to the specific optimization problem at hand is pivotal for attaining optimal performance. Given GA's widespread recognition, we won't delve deeply into its components within this article. For a comprehensive understanding, readers are directed to established research [11, 36]. However, it's essential to tailor the algorithm to the unique contours of the problem in focus. Subsequent sections will elucidate how we've molded the conventional GA to align with our distinct context.

**4. Results and Analysis**

This section presents the results and analysis of this study. A through summary of the previous findings of this research project [8] and identified causes of agile-quantum projects are discussed in sub-section 4.1. The results and analysis of developed success probability prediction model is detailed in section 4.2.

**4.1 Previous study and identified causes-practices**
We extended the previous empirical study [8] findings by summarizing the core categorizing of identified challenges with respect to their relevant causes and practices. Thematically, we categorized the identified challenges across core four categories- knowledge and awareness, quantum-aware tools and technologies, sustainable scaling, and standards and specifications.
We now summarize these categories with respect to the causes of the challenges and their respective best practices.

**Knowledge and awareness**: This category encapsulates the substantial challenges related to the knowledge and understanding of using traditional agile practices within the quantum software development realm. The most prominent challenges in this category include *knowledge gap*, *team dynamics*, and *quantum software development education*. These challenges stem from a lack of *domain specific knowledge, lack of market interest, limited academic research, funding sources*, and *gap between research and practice.* To tackle these challenges, interviewees suggested a list of best practices that recommend fostering team unity through team-building and cohesiveness-enhancing activities, and improving quantum literacy via targeted workshops and training. Strengthening the infrastructure and promoting industry awareness, alongside academic research support, can advance the field. Similarly, understanding the commercial and operational complexities of quantum systems and applying domain best practices are also crucial for effective agile integration. This analysis underscores the criticality of addressing the "Knowledge and Awareness" barriers to successfully scale agile practices in this cutting-edge domain. The list of identified challenges, their causes and best practices is provided in Appendix A.

**Sustainable scaling**: This category explores the challenges of bringing agile practices sustainably within the quantum software development sphere, highlighting two principal challenges: *ethically aligned quantum software development* and *agile-quantum ecosystem*. Their underlying causes—*complex technical requirements*, *cross-disciplinary integration difficulties, emerging working culture, quantum-agile interface stability, security, reliability,* and *rapid pace of innovation*—demand targeted practices for resolution. In response, interviewees suggested strengthening quantum domain-specific expertise and establishing defined knowledge management criteria as fundamental steps. Building a technical infrastructure dedicated to quantum software development activities is also important. The development of universally applicable quantum-enabled simulators could democratize access to quantum technologies across various platforms. Furthermore, leveraging software to reduce energy consumption through 'green agile' techniques could contribute to more sustainable quantum computing practices. Tools for continuous quantum-classical hybridization are essential to bridge the gap between traditional/classical computing infrastructure and quantum systems. Finally, positioning quantum ethics at the nexus of quantum information science, technology ethics, and moral philosophy is necessary for a holistic assessment of the impacts posed by emergent quantum technologies. These practices aim to create a scalable, ethical, and agile framework that is responsive to the unique demands of quantum software development.

**Quantum-aware tools and technologies:** This category of challenges emerged from two major challenging factors- *classic-quantum tailoring* and *continuous SE infrastructure*. The core causes of these challenges include *right tool for the right job*, *lack of industrial interest*, *limited resources*, *transforming process from legacy to quantum software*, and *technological paradigm shift*. To navigate these challenges, a suite of best practices is essential. Developing agile toolkits specifically for quantum software will facilitate the precise tailoring needed for quantum tasks. Regular evaluation and selection of development tools for agile based quantum software development will ensure a consistent alignment with project requirements. Allocating dedicated resources within agile frameworks to quantum exploration can address the limitations of funding and focus. Forming quantum interest groups within agile teams can spark industrial interest and foster collaborations. Designating sprints for the transformation from legacy to quantum software allows for a structured approach to this complex transition. Workshops aimed at enhancing agility in quantum computing can prepare teams for the technological paradigm shift. Finally, advancing CI/CD pipelines to accommodate quantum development ensures that the agile practices can be sustained in a continuous SE infrastructure. Integrating these practices can provide a robust framework for overcoming the challenges of agile-quantum software development realm.

**Standards and specifications:** In the domain of standards and specifications for agile-based quantum software development, there are two pivotal challenges: *process standardization* and *optimum documentation*. These challenges stem from *varying interpretations of agile methodologies*, *rapid evolution of quantum technologies*, *complexity of quantum computing concepts*, the *necessity to integrate with existing standards*, and *interdisciplinary collaboration barriers*. Addressing these challenges necessitates a coherent set of practices:

For process standardization, maturity models tailored to agile-quantum development are essential. They assess and guide teams toward refined agile practices. Complementing these models, standard assessment protocols are necessary to gauge the effectiveness of agile practices and documentation in the quantum software development environment. A comprehensive integration framework is also necessary, providing best practices for blending agile methodologies with the quantum software development process, inclusive of standardized process checklists and templates. Simultaneously, establishing quality standards for documentation specific to the quantum software domain is essential. These standards should guide teams in creating documentation that is clear, concise, and essential for the project at hand. Agile teams should incorporate sprints dedicated to ensuring compliance with established frameworks and standards, allowing for consistent realignment with best practices throughout the development cycle. Furthermore, advocating for the utilization of Software Development Kits (SDKs) with inherent standards and documentation guidelines can streamline adherence to best practices, reducing the overhead for agile teams. Developing agile-quantum playbooks offers a standardized yet adaptable approach to procedures and documentation templates across various quantum development projects. Furthermore, to enhance documentation quality, conducting workshops is key, where agile teams are trained to craft documentation tailored to the unique demands of quantum software development, emphasizing the principles of agility and relevance. Likewise, dedicating specific sprints to the iterative development and refinement of documentation ensures that it evolves iteratively with the rapid development cycles characteristic of quantum software projects. This proactive approach ensures the documentation remains current with the latest project insights, architectural decisions, and code modifications, achieving an optimal balance throughout the project's lifecycle.

**4.2 Agile-Quantum Software Project Success Prediction Model (AQSSPM)**
This section provides an overview of the survey participants' demographics (see section 4.2.1), describe the application process of Genetic Algorithms (GA) (see section 4.2.2), and the results of the proposed model are discussed in section 4.2.3.

**4.2.1 Experts Demographics**
We conducted a frequency analysis to examine the collected data, a method suitable for evaluating multiple variables, including numeric and ordinal data. Our study involved 104 participants from 16 countries across four continents, covering 12 professional roles and 12 different project directions, as shown in Figure 1(a-c). A significant number of participants have experience levels between 3-5 years in their fields, as depicted in Figure 1(d). We also explored gender representation in the collected responses. Participants were asked to self-identify their gender. The data revealed a majority of male participants at 64%, compared to 19% females. 21% opted not to specify their gender, as shown in Figure 1(e).

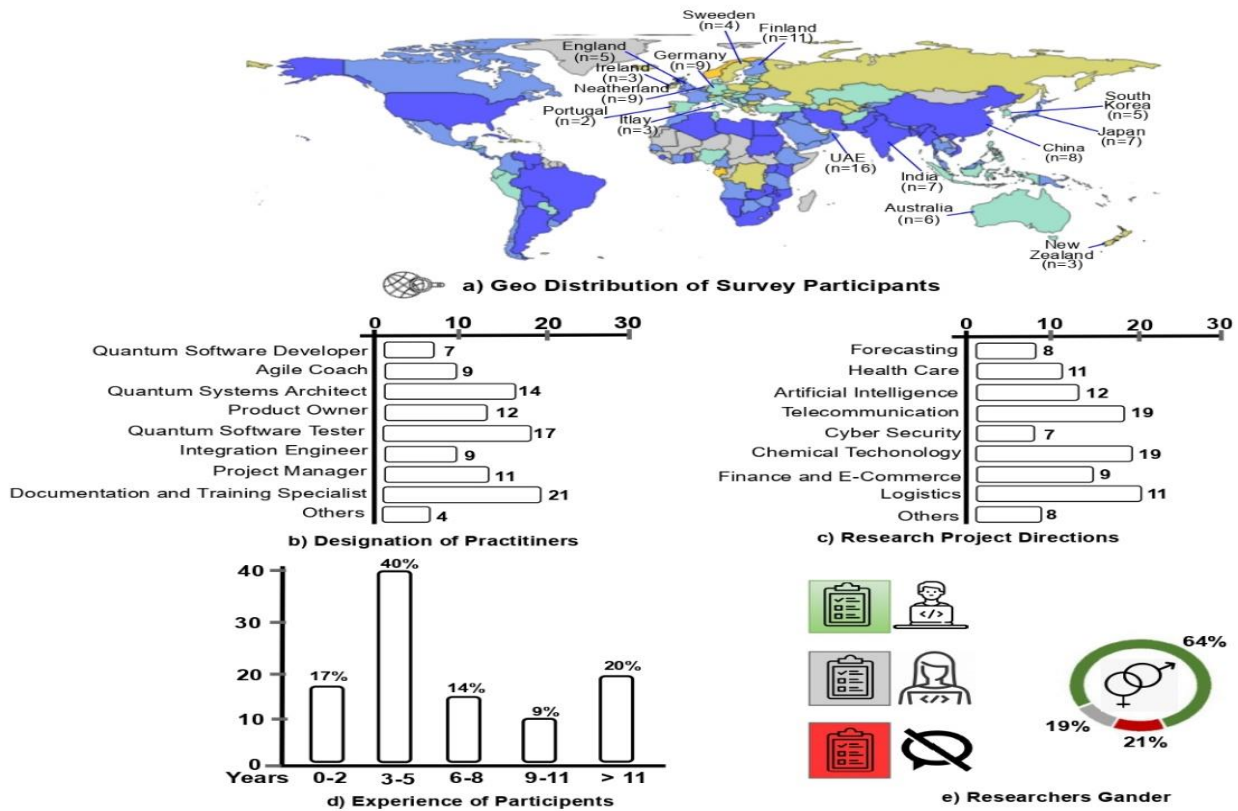

**Figure 1:** Demographics data collection process participant

To understand our participants' professional roles, we used thematic mapping. This helped categorize their roles into 9 different categories, shown in Figure 2(b). The roles of documentation and training specialist were the most common, with 21 participants. Additionally, we identified 9 main working domains among the participants. Telecommunication and chemical technology emerged as dominant sectors, each with 19 participants, as illustrated in Figure 1(c).

### 4.2.2 Genetic Algorithm Application Process

This section formally presents the optimization problem of success probability maximization in the presence of cost. It is explained that the GA algorithm needs to be customized for this specific problem. Additionally, the necessary modifications to GA and its components are also discussed. We have employed the GA in the following steps:

**Step-1 Representation**

In the current study, each potential solution, i.e., chromosome, comprises an ordered set of values. The order of the values in the set corresponds to the order of attributes described in Figure 2. The first value in the set represents the first attribute scale, the second value represents the second attribute level, and so on.

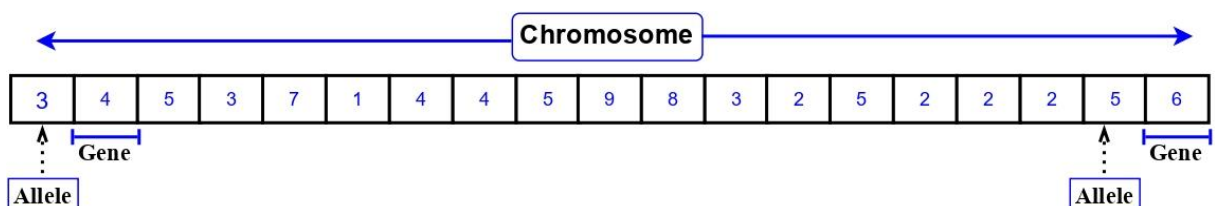

Figure 2: The structure used to represent a chromosome

**Step- 2 Initialization**

The research utilized a sample size of 50, where the initial set was produced randomly. Precisely, 50 chromosomes were established at the onset of the algorithm, and every chromosome component was allocated a random whole number ranging from 1 to 9, inclusive.

**Step-3 Fitness function**
The efficiency of a chromosome is determined using Equation 15, denoting the chromosome's fitness metric. A superior fitness metric signifies a heightened likelihood of success and reduced related expenses.

**Step -4 Constraint handling**
Numerous practical challenges come with constraints, suggesting that a solution might not always be viable. Chromosomes formed in GA might sometimes reside in an impractical search area, indicating they don't adhere to the parameters outlined in Equation 15. To ensure that chromosomes conform to the stipulated guidelines, they are designed to meet specific constraints. This ensures they remain within boundaries, even after iterative changes across generations. The value range was confined between $S_{max}$ and $S_{min}$, and the GA settings were tailored to produce chromosomes within these acceptable zones.

**Step-5 Selection & Reproduction**
The present research adopted the roulette-wheel selection method from the given options due to its straightforward nature and efficiency.

**Step–6 Crossover**
A set probability value of 0.8 is employed for chromosome crossover, and the single-point crossover technique was chosen for its straightforward approach.

**Step–7 Mutation**
We opted for the random mutation approach, which has a probability of 0.1. This means there is a 10% chance for each chromosome to be substituted with a random number ranging between $S_{max}$ and $S_{min}$.

**Step-8 Stopping criteria.**
The study adopted a threshold of 100 generations as the termination condition. All essential elements and their corresponding values pertinent to the optimization issue outlined in Equation 15 can be found in Table 2.

Table 2: The values of the GA parameters

| Parameter Name | Value |
| --- | --- |
| Max. iterations | 100 |
| Size of population | 50 |
| Total genes | 40 |
| Type of gene | Integer |
| Max. gene value | 9 |
| Min. gene value | 1 |
| Selection method | Roulette wheel |
| Crossover method | Single point |
| Probability of crossover | 0.8 |
| Mutation method | Random |
| Probability of mutation | 0.1 |
| Fitness function | Efficacy (Equation 15) |

**Execution of GA-** In the sections above, we elaborated on all the essential components and their associated values required for the GA execution. Algorithm 2 offers a thorough recapitulation of the algorithm's steps.

---

**Algorithm 2:** Steps of Genetic Algorithm employed

**Step 1:** Train a predictive model to produce success probability.
**Step 2:** Set the values of GA parameters.
**Step 3:** Randomly generate the initial population of chromosomes, says *Pop*.
**Step 4:** Calculate the success probability *Prob*(*S*) of each chromosome *S* in Pop using Equation 8 with the help of the predictive model mentioned in Step 1.
**Step 5:** Calculate the normalized cost of each chromosome *S* in Pop using Equation 13.
**Step 6:** Calculate efficacy *E*(*S*) for each chromosome *S* in Pop using Equation 15, and it will serve as the fitness function *f*(*S*) of S.
**Step 7:** Generate child population C by applying selection, crossover, and mutation operators on Pop.
**Step 8:** Pop ← C.
**Step 9:** Repeat Steps (4-8) until the stopping criteria are met.

---

Upon successful execution of Algorithm 2, the chromosome labeled SB, having the optimal fitness value, is retained. Leveraging the scale values of each variable encompassed in SB, an optimal solution conducive to project success can be derived.

### 4.3.3 AQSSPM results

This section presents the final results obtained by implementing the GA with Naive Bayes classifier (NBC) and the Logistic regression (LR) model. Furthermore, the results of NBC and LR correlation analysis are also presented in sub-sequent section.

- **Results with GA-NBC**

The results of the GA-NBC for agile-quantum project success probability are presented in Table 3 while the best fitness order of the identified 19 causes are presented in Table 4. Initially, the success probability and cost were 53.17% and 0.463, respectively. After 100 generations, the optimal solution yielded a cost of 0.403 and a success probability of 99.68%, signifying a 46.51% improvement in success probability and 6% decrease in cost, as given in Table 3. The results from the NBC and GA analyses indicate the initial cost of developing quantum software using agile was relatively high.

The initial low success rate and high cost in quantum software development using agile methodologies can be attributed to various causes faced by the industry as given in Appendix A. One of the primary obstacles is the C1 (lack of domain-specific knowledge), which hampers the efficient application and understanding of quantum principles. Furthermore, a noticeable C2 (lack of market interest), combined with C3 (limited academic research), means there is a dearth of foundational and advanced knowledge being disseminated and applied. Furthermore, C4 (funding sources) are limited, making it challenging to invest in innovative solutions and tools that could streamline the development process. The frequent iterations underlying to agile methods exacerbate resource consumption, potentially straining already limited resources. C6 (cross-disciplinary integration difficulties) adds another layer of complexity, and developing quantum code at the gate level remains a specialized and intricate task.

The agile approach demands the C8 (right tool for the right job), but this is often lacking in the agile-quantum realm. The stability, security, and reliability of the quantum-agile interface are still under scrutiny, and the dynamics and evolution of business models in this nascent field can be unpredictable.

Classical agile practices might not seamlessly integrate with quantum software systems, raising concerns about their compatibility.

Moreover, the quantum software development life cycle is still in its infancy, with many organizations grappling with the transition from classical to quantum software. This transformation is further hampered by a discernible gap between research and practical application. As the industry evolves, C18 (emerging working cultures) and the imperative of adhering to business ethics and corporate governance can influence the trajectory of agile adoption for quantum software development.

However, as the process matures and these causes are addressed, there is potential for a significant increase in success probability, as endorsed by the GA-NBC analysis. The subsequent decrease in cost also suggests that with time and focused efforts, the quantum software industry can optimize its agile methodologies, leading to more efficient and cost-effective outcomes.

In the realm of agile-quantum software development, several causes could influence the success probability and cost. The results given in Table 4 suggest the best fitness of identified causes to increase the project success probability and to decrease the cost. For instance, the cause C6 (Cross-disciplinary integration difficulties) holds the most significant impact. This highlights the need of integrating cross-disciplinary practices within quantum projects to achieve optimal outcomes. Following closely are causes C10 (limited resources) and C11 (varying interpretations of agile methodologies). The scarcity of specialized tools and expertise underscores the importance of resource optimization. Additionally, the interpretations of agile methods and techniques should be rigorous, emphasizing stable protocols and efficient integrations. Together, addressing these primary causes can significantly bolster the success rate of agile-quantum projects.

Table 3. The initial and ending fitness achieved by the NBC model for all variables

| Stages | Generations | Initial success probability | Ending success probability | Change in probability | Initial Cost | Ending Cost | Change in cost |
|---|---|---|---|---|---|---|---|
| Naive–bayes | 100 | 53.17% | 99.68% | +46.51% | 0.463 | 0.403 | -6.0% |

Table 4. The best fitness of causes obtained from all stages of GA-NBC

| Model | C1 | C2 | C3 | C4 | C5 | C6 | C7 | C8 | C9 | C10 | C11 | C12 | C13 | C14 | C15 | C16 | C17 | C18 | C19 |
|---|---|---|---|---|---|---|---|---|---|---|---|---|---|---|---|---|---|---|---|
| Naive–bayes | 7 | 7 | 2 | 5 | 4 | 9 | 2 | 6 | 6 | 8 | 8 | 4 | 6 | 5 | 3 | 2 | 3 | 6 | 2 |

- **Results with Logistic Regression Model**

The success probability of agile-quantum projects was also assessed using a GA-LR model. Table 5 presented that when the LR model is paired with GA, it considerably impacts the project's outcome. Initially, the project had a success rate of 55.52% at a cost of 0.496. But after 100 iterations with the GA-LR, the outcome improved dramatically, showing a success rate of 98.99% at a lowered cost of 0.409. This indicates a notable 43.47% improvement in success probability and an 8.70% cost reduction (Table 5). This indicated that by focusing on the main causes, we can increase the agile-quantum project's success probability by 43.47% while also cutting down costs.

Therefore, we determined the best fitness of identified causes to increase the project success probability and to decrease the cost by applying GA-LR model (Table 6). For instance, C1 (lack of domain specific knowledge) stands out as the most critical cause, as the field of quantum software engineering demands specialized knowledge for effective project execution. Without a deep understanding of the domain, teams can face significant barriers, leading to inefficiencies and increased costs. Moreover, C10 (limited resource), agile methodologies, known for their iterative nature, require consistent and specialized resources. A scarcity can hinder the smooth progression of the development cycle, potentially escalating costs. C11 (varying interpretations of agile methodologies) is equally crucial. As quantum computing introduces new paradigms, ensuring a seamless, secure interface with agile practices becomes paramount. Any vulnerabilities can not only risk the project's success but also inflate costs due to unforeseen issues. Additionally, C2 (lack of market interest) plays a pivotal role. For any project to be successful, it needs to align with market demands. If the agile-quantum project doesn't resonate with market interest, its chances of success diminish, and resources invested might not yield the desired return. Lastly, C6 (cross-disciplinary integration difficulties) is essential to the success of agile-quantum projects, as it enables the harmonious combination of diverse expertise and insights. Successfully addressing C6 is key to fostering innovation and ensuring the efficacy of agile-quantum initiatives. In essence, prioritizing and addressing these causes head-on can significantly enhance the success trajectory of agile-quantum projects while ensuring cost-effectiveness.

Table 5. The initial and ending fitness achieved by the LR model for all variables

| Stages | Generations | Initial success probability | Ending success probability | Change in probability | Initial Cost | Ending Cost | Change in cost |
|---|---|---|---|---|---|---|---|
| Logistic–regression | 100 | 55.52% | 98.99% | +43.47% | 0.496 | 0.409 | -8.70% |

Table 6. The best variables are obtained for all the stages from the LR classifier

| Model | C1 | C2 | C3 | C4 | C5 | C6 | C7 | C8 | C9 | C10 | C11 | C12 | C13 | C14 | C15 | C16 | C17 | C18 | C19 |
|---|---|---|---|---|---|---|---|---|---|---|---|---|---|---|---|---|---|---|---|
| Logistic–regression | 8 | 7 | 2 | 5 | 4 | 7 | 2 | 6 | 6 | 8 | 8 | 3 | 5 | 5 | 4 | 1 | 4 | 7 | 2 |

- **Comparison Between NBC and LR Based Best Fitness of Identified Causes**

In this section, we compared the results (best fitness of causes) derived from the GA-NBC with those of the GA-LR approach to gain a holistic insight into their respective efficacies.

*Variable Best Fitness Order* - Our proposed framework for gauging the success probability in agile-quantum projects presents two key benefits. Firstly, it suggests an ideal cost allocation across different causes to amplify the success probability of agile-quantum project. Secondly, it offers insights into the comparative significance of the identified causes.

Table 3 presents the initial and ending cost to fix the identified causes necessary to approach a near-perfect project success rate. The results suggests that some causes might not be pivotal, as elevated success percentages can be maintained even with reduced cost allocations. For instance, cause C7 (rapid pace of innovation) need 2 units to attain a success probability of 98.68% using the GA-NBC. Conversely, C6 (cross-disciplinary integration difficulties) requires an allocation of 9 units, indicating a more substantial financial and effort commitment to optimize the probability of success in the agile-quantum project (see Table 4).

Further assessments employing GA-LR learned the financial allocation per cause critical for enhancing the success probability of agile-quantum project. Table 5 reveals that certain causes employ minimal influence on success, sustaining high success percentages even with minimum allocated funds and effort. For instance, C3 (limited academic research) demands only a 2-unit allocation, while C1 (lack of domain-specific knowledge), C10 (limited resources), and C11 (varying interpretations of agile methodologies) each necessitate 8 units. Based on these findings, emphasizing C1, C10, and C11 while limiting the budget for C3 ensures an elevated success probability i.e., 99.29%.

By analyzing the relationship between each cause financial allocation and the project's success probability, we can order the causes based on their cost-effectiveness and overall relevance. Employing results from the GA-NBC (refer to Table 4) and GA-LR (see Table 6), we allocate rankings to each cause and the determined ranks are given in Table 7.

For example, C6 (cross-disciplinary integration difficulties), with an allocation of 9 units in the NBC analysis and 7 units in the LR analysis, holds the first and second ranks, respectively, highlighting its paramount importance. Using a similar methodology for all causes, we compile a comprehensive ranking (displayed in Table 7), assisting professionals in emphasizing the pivotal cause for agile-quantum projects, thereby substantially improving the project success probability.

Table 7: Best fitness ranking of causes for both classifiers

| Model | Particulars | C1 | C2 | C3 | C4 | C5 | C6 | C7 | C8 | C9 | C10 | C11 | C12 | C13 | C14 | C15 | C16 | C17 | C18 | C19 |
|---|---|---|---|---|---|---|---|---|---|---|---|---|---|---|---|---|---|---|---|---|
| **GA-NBC** | Best Fitness | 7 | 7 | 2 | 5 | 4 | 9 | 2 | 6 | 6 | 8 | 8 | 4 | 6 | 5 | 3 | 2 | 3 | 6 | 2 |
| | Ranks | 3 | 3 | 8 | 5 | 6 | 1 | 8 | 4 | 4 | 2 | 2 | 6 | 4 | 5 | 7 | 8 | 7 | 4 | 8 |
| **GA-LR** | Best Fitness | 8 | 7 | 2 | 5 | 4 | 7 | 2 | 6 | 6 | 8 | 8 | 3 | 5 | 5 | 4 | 1 | 4 | 7 | 2 |
| | Ranks | 1 | 2 | 7 | 4 | 5 | 2 | 7 | 3 | 3 | 1 | 1 | 6 | 4 | 4 | 5 | 9 | 5 | 2 | 7 |

*NBC and LR Correlation Analysis* – We undertook a non-parametric statistical analysis to identify the similarities and differences between the optimal fitness rankings derived from the Naive Bayes Classifier (NBC) and Logistic Regression (LR) [37, 38]. Several relative research studies have adopted this non-parametric statistical analysis approach [23, 39, 40]. For a uniform metric, we employed Spearman's Rank-Order Correlation Coefficient, organizing each factor associated with the agile-quantum project in accordance with the supreme fitness values yielded by both NBC and LR (see Table 7). This correlation coefficient gauges the linear association between factors, with potential values spanning from +1 to -1. A coefficient of +1 signifies an absolute linear correlation.

A Spearman's Rank-order correlation coefficient value was ($r_s=0.955$, $p=0.000$), as presented in Table 8. This high correlation value underscores a strong association between the rankings derived from Naive Bayes Classifier (NBC) and Logistic Regression (LR) methodologies. This remarkable agreement implies that practitioners can utilize either of these techniques to pinpoint the significant factors (causes) affecting agile-quantum projects.

Although there is significant alignment, minor variations in rankings attributed to specific factors by NBC and LR were detected. For example, C1 (lack of domain-specific knowledge) secured the 3rd rank using NBC but ascended to the top rank with LR. Similarly, C2 (lack of market interest) and C4 (funding resources) showed slight shifts in rankings (as seen in Table 8). These nuanced differences hint at a measure of variation in the ranking system between the two techniques, a factor to bear in mind when discerning pivotal aspects of agile-quantum projects.

Beyond the Spearman's Rank-order correlation, an independent t-test was applied to evaluate the average rank differences generated by both methods (see Table 9). The Levene's Test outcome was non-significant ($p=0.967>0.05$), confirming uniform variances, and our analysis proceeded under this assumption. The t-test results ($t = 1.195$, $p = 0.240 > 0.05$) revealed no significant differences between the ranks produced by both methodologies. This uniformity confirms that, for determining variable importance, practitioners can confidently adopt either NBC or LR, optimizing resource distribution and amplifying the success prospects of agile-quantum projects.

Table 8: Correlation between the ranking of NBC and LR

|  |  |  | LR_Ranking | NBC_Ranking |
|---|---|---|---|---|
| Spearman's rho | LR_Ranking | Correlation Coefficient | 1.000 | 0.955** |
|  |  | Sig. (2-tailed) | . | 0.000 |
|  |  | N | 19 | 19 |
|  | NBC_Ranking | Correlation Coefficient | 0.955** | 1.000 |
|  |  | Sig. (2-tailed) | 0.000 | . |
|  |  | N | 19 | 19 |
| **. Correlation is significant at the 0.01 level (2-tailed). | | | | |

Table 9: Independent Samples t-test of NBC and LR rankings

|  |  | Levene's Test for Equality of Variances | | t-test for Equality of Means | | | | | | |
|---|---|---|---|---|---|---|---|---|---|---|
|  |  | F | Sig. | t | df | Sig. (2-tailed) | Mean Difference | Std. Error Difference | 95% Confidence Interval of the Difference | |
|  |  |  |  |  |  |  |  |  | Lower | Upper |
| Rank | Equal variances assumed | 0.002 | 0.967 | 1.195 | 36 | 0.240 | 0.89474 | 0.74886 | -0.62402 | 2.41349 |
|  | Equal variances not assumed |  |  | 1.195 | 35.941 | 0.240 | 0.89474 | 0.74886 | -0.62410 | 2.41358 |

## 5. Summary of results

Now we present the summary of the developed AQSSPM model results in section 5.1 and the core implications of this research project are reported in section 5.2.

### 5.1 AQSSPM

The objective of this research project was to explore the challenges, causes behind these challenges and the most effective practices to address them. Finally, based on the identified causes, develop the AQSSPM model which could evaluate the success probability of agile-quantum projects, serving as a roadmap to industry practitioners for informed decision-making. Central to this model's development is a thorough examination of key causes, validated by industry experts through structured questionnaire study. To develop AQSSPM, we used the GA, NBC and LR approaches, which analyze the causes in terms of financial allocation, and cost (team composition, monetary pledges, and other essential assets). Following this, our study gives deep insights of the core causes need to be addressed on priority basis to increase the agile-quantum project success and to minimize the cost.

The developed AQSSPM is presented in Figure 3, that shows that using GA alongside NBC, we observed the success probability of an agile-quantum project soaring from an initial 53.17% to a remarkable 99.68% over its lifecycle, while concurrently reducing costs from 0.463% to 0.403%. On other hand the GA with LR, the success rate leaped from 55.52% to 98.99%, coupled with a cost saving, as costs declined from 0.496% to 0.409% after 100 iterations.

Enhancing the power of GA in combination with both NBC and LR granted us the ability to perceive the fundamental determinants behind the success of agile-quantum projects. This evaluation gives a body of knowledge to practitioners with a solid roadmap, underscoring the significance of domain specific knowledge, proper cross-disciplinary integration, availability of resources, adaptability to technological paradigm shifting, and increase market interest. Acknowledging and ranking these causes can revolutionize project execution, setting the stage for agile-quantum endeavors that are successful and financially practical. In summation, the result of this study serves the industry practitioners and researchers with the knowledge and tool required to amplify the success probability of their agile-quantum projects.

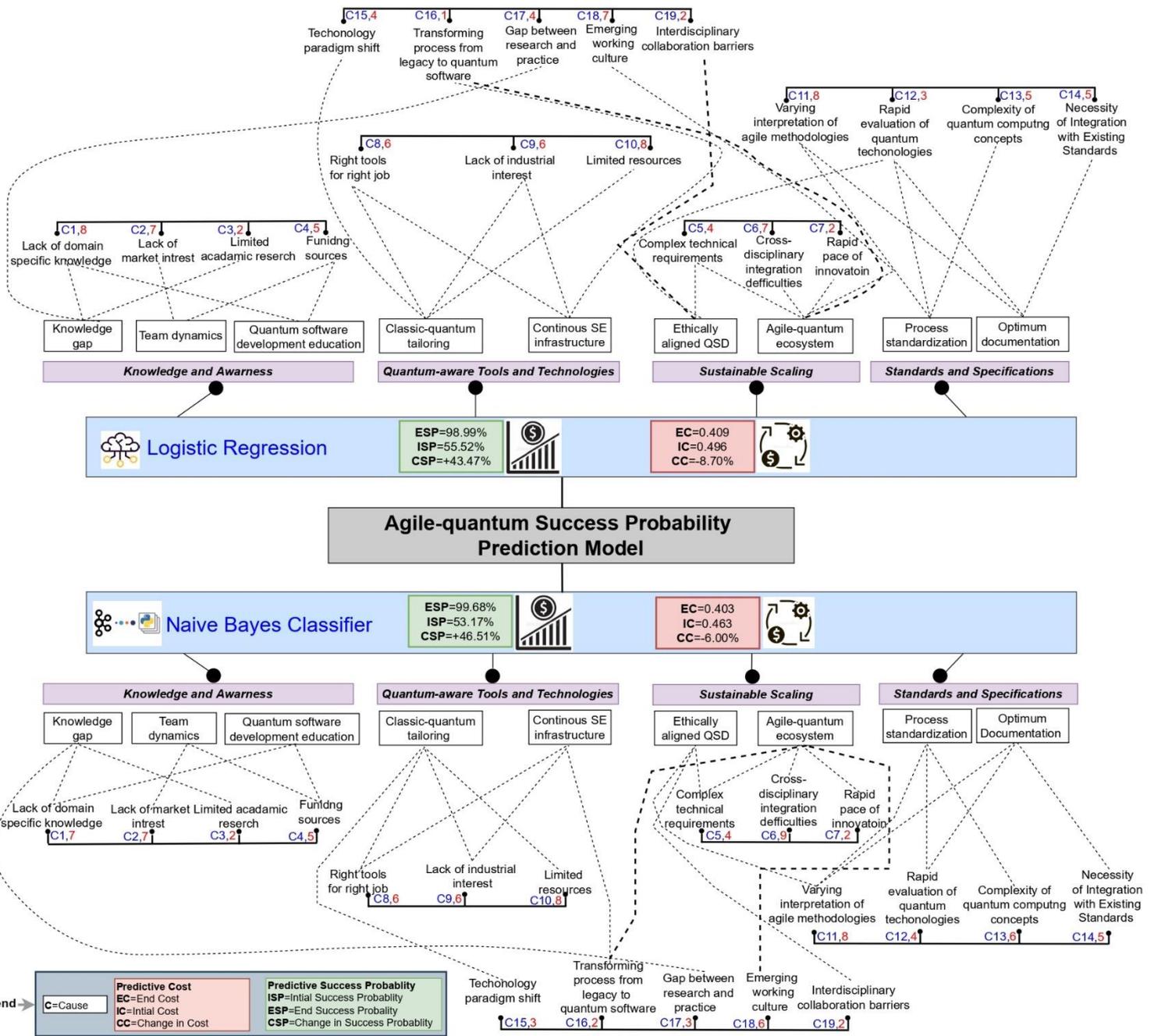

**Figure 3**: Agile-quantum project success probability prediction model

## 5.2 Academic Implications

The research presented in this paper has substantial academic implications, particularly for the growing field of QSE. By investigating the suitability of traditional agile methodologies within quantum software development, this study contributes to the theoretical foundation of QSE, an area that is still in its infancy. The paper extends the knowledge base by identifying challenges specific to quantum software development, their root causes, and offering best practices to overcome them, thus filling a gap in the existing literature.

The introduction of a predictive model based on genetic algorithms represents an innovative approach to address decision-making in quantum software projects. This model, which employs nature-inspired algorithms to evaluate project success probability, not only adds a new tool for academic research but also sets the stage for future studies to refine and validate predictive models in this domain. It encourages further empirical research to explore the effectiveness of agile practices in quantum software development, potentially leading to the development of new frameworks and methodologies that are specifically designed for the quantum realm.

Moreover, the interdisciplinary nature of this study, bridging software engineering with QC, illustrates the importance of cross-disciplinary research in addressing complex technological challenges. It provides a template for how agile methodologies can be adapted to fit new and emerging fields, offering a blueprint that can be studied and replicated across different domains of technology research.

Based on this study, future research could be conducted to integrate traditional software engineering methodologies, tools, and techniques within the quantum software domain. This direction would be critical, considering that the current infrastructure for QSE is emerging and largely unexplored. Subsequent studies could evaluate how existing software engineering practices could be adapted for quantum computing, thus maximizing the concept of hybridization between conventional and quantum software development. This integration is essential for establishing a robust quantum software engineering ecosystem that can support the rapid evolution and unique demands of quantum computing technologies.

### 5.3 Industrial Implications

For the industry, the implications of this research are equally significant. The transition from traditional computing to quantum computing represents a paradigm shift in how computational tasks are approached and executed. By offering a detailed analysis of the challenges and best practices for adopting agile methods in quantum software development, the paper provides industry professionals with a framework to navigate this transition effectively.

The predictive model introduced in this paper serves as a practical tool for organizations considering the integration of existing agile practices into their quantum computing projects. It enables them to make informed decisions about the potential costs and outcomes of their projects, thereby reducing uncertainty and increasing the likelihood of successful implementation.

Additionally, by highlighting the need for quantum-aware tools and technologies, the research underscores the market demand for specialized solutions that organize to the unique requirements of quantum software development. This could stimulate investment and innovation in the development of such tools, fostering a supportive ecosystem for the growth of the quantum software industry.

Overall, the findings from this research provide valuable insights that can guide the development of quantum software, ensuring that the industry can keep pace with technological advancements while maintaining software quality, sustainability, and ethical standards.

### 6 Threats to Validity

Various potential threats could affect the validity of the study findings. We have examined potential threats across the four essential categories of validity threats: internal validity, external validity, construct validity, and conclusion validity, as defined by Wohlin et al.[41]

## 6.1 Internal Validity Threats:

Internal validity refers to the extent to which particular factors affect the results and findings of the collected data. The internal validity of our study may be compromised by several factors based on the survey design and the subjectivity of responses. The use of a closed-ended questionnaire, while beneficial for standardizing responses, may limit the depth of insight into the complex integration of traditional and quantum software engineering practices. The 9-point rating scale assumes a linear perception of the attributes (variables) being measured, which may not accurately capture the detailed opinions of agile and quantum computing practitioners. Additionally, the study validity can be compromised if participant responses are influenced by their individual levels of familiarity and hands-on experience with quantum software development, potentially introducing biases that could skew the results. Moreover, the misinterpretation of survey questions by the participants could jeopardize the accuracy of the data collected. To minimize these threats, we piloted our survey instrument, refining it with feedback from experts in empirical software engineering to enhance its reliability, understandability, structure, and validity.

Furthermore, the potential biases stemming from the use of genetic algorithm, NBC, and LG—such as premature convergence, feature independence assumption, and assumption of linearity—are primarily threats to the internal validity of the study findings. These biases can influence the causal relationships and the accuracy of the model predictions due to selection bias, and assumptions about data characteristics that do not hold true. However, we collected data from 104 participants across 16 different countries, consisting of 12 professional roles and project directions which significantly enhances the diversity and representativeness of the study sample. This wide range of data helps mitigate biases related to sample selection and algorithm assumptions by ensuring a broad spectrum of insights and experiences. It contributes to the model generalizability and reduces the risk of overfitting.

## 6.2 External Validity Threats:

External validity concerns the extent to which a study results can be generalized beyond the particular sample, setting, or conditions examined. This involves considering how the study findings are applicable to other groups, environments, or scenarios not directly investigated in the research. The external validity of our study is based on the representativeness of our participant pool. With data collected from 104 agile and quantum computing practitioners, there is a risk that our sample may not be sufficiently diverse or large enough to generalize findings across the entire field of quantum software development. Despite efforts to ensure a diverse population, the possibility that our sample is not fully representative of the global community or all expertise levels in quantum computing and agile methodologies remains a threat to the validity of our study findings. Furthermore, the rapid evolution of quantum computing technology may mean that our findings have a limited shelf-life, potentially reducing their applicability as the field advances. Therefore, to mitigate the stated threats, our survey reached participants from 16 countries across four continents, broadening the cultural and geographical scope of our study. Our sample size is on par with similar studies in the growing field of quantum software development [6], supporting the generalizability of our results. The international reach of our survey supports a global perspective, potentially reflecting wider trends in quantum computing practices.

Additionally, the broader applicability of the proposed model presents a significant threat to the external validity of the study findings. Its practical use in real-world projects could be affected by certain factors which were not directly considered on this study, such as existing (classical) software development practices and the software practitioners willingness to adopt new technologies. To address this issue and improve the study external validity, future research should aim at conducting case studies or pilot implementations of the model in real-world quantum software development projects.

### 6.3 Construct Validity Threats:

Construct validity addresses the degree to which the survey questions accurately measure the concepts they are intended to assess. In our study, the construct validity is dependent upon how well the survey questionnaire design captures the study aim of integrating traditional agile practices for developing a quantum software system. There is a possibility that the survey items might not fully encapsulate the challenges of bringing agile practices into a quantum software environment, which could result in an incomplete understanding of the targeted problem. To ensure that the constructs are valid and the survey questions are well-calibrated, we incorporated iterative feedback from a diverse range of experts in the field to fine-tune the survey items, thereby enhancing the relevance and accuracy of the constructs being measured.

### 6.4 Conclusion Validity Threats:

Conclusion Validity refers to the level of credibility or legitimacy of the study conclusions drawn from the overall study results. The conclusion validity of our study is dependent on the questionnaire design, which was based on previous studies and literature, there is a risk that certain pertinent variables may have been overlooked, which could lead to incorrect inferences. The use of a 9-point scale and the subjective nature of the responses pose a risk of misclassification or overestimation of the strength of associations. Moreover, with a sample size of 104 participants, our study might not have the power to detect subtle but meaningful differences or trends, which could lead to either Type I or Type II errors in our conclusions. To minimize the conclusion validity threats, the survey instrument was thoroughly reviewed by subject matter experts for clarity and coherence and was structured in alignment with established empirical software engineering study designs [23, 42].

## 7 Related Work

We now summarized the related work focusing on quantum software development with respect to software processes (section 5.1) and probabilistic models developed using the genetic-based algorithm (section 5.2).

### 7.1 Quantum Software Development

Quantum software development processes involve designing, implementing, and testing software that runs on QC platforms. This process requires an understanding of quantum algorithms, quantum logic gates, and the unique behavior of quantum bits (qubits). Based on the given characteristics of QC, its challenging to develop quantum-specific software systems using the traditional processes. Few studies have been conducted that focused on defining quantum software processes, methods, and frameworks.

For instance, González[43] described that in the realm of QC, hardware advancements, characterized by the rapid increase in qubit numbers, maximize the need for the core approaches used to execute the sophisticated algorithms and develop commercial systems. However, programming and engineering methodologies have not kept pace, lacking the specific approaches required for the novel challenges and risks inherent to quantum technology. This paper responds to the need for advanced project management strategies in quantum software development, advocating for a hybrid framework that orchestrates traditional agile methods with the unique aspects of quantum programming. This proposed framework aims to organize the complexities of managing projects that blend classical and quantum development, ensuring readiness for the future landscape of software engineering in the quantum era.

Khan et al.[13] presented an idea paper and discussed that in realm of HOC and QC, where resource usage is costly, software development demands a nearly flawless product before execution. This paper proposes an agile development methodology tailored for quantum software, where initial stages use interactive, low-cost computations for conceptualization and experimentation. As the project matures, more resource-intensive tasks like algorithm optimization are batch-processed, integrating iterative development with the resource-conscious demands of QC.

Weder et al.[16] reported that the complexity of developing quantum software applications requires multidisciplinary expertise, which boosts the emergence of quantum software engineering as a field focused on the application of systematic software principles. Hybrid quantum applications, integrating quantum and classical components, need life cycles that address both, plus the orchestration of execution and data exchange. This study introduces a comprehensive quantum software development life cycle, examining phases, tools, and the integration of different software artifacts life cycles. This article outlines hybrid quantum application fundamentals, presents the proposed life cycle, discusses assumptions and limitations, reviews related work, and concludes with future perspectives.

Khan et al.[8] explore the organization of agile development practices within quantum software development through encapsulating the opinions of practitioners from 10 countries. Findings suggest agile practices are applicable for developing quantum software but also uncover unique challenges in integrating these agile approaches effectively. The research offers insights for adapting agile practices to the specific demands of quantum software engineering, paving the way for the advancement of future quantum software systems and applications.

### 7. 2 Genetic Algorithm-Based Probabilistic Models

A genetic algorithm is a search heuristic that uses natural selection and genetic processes such as mutation and crossover, evolving a population of solutions to optimize results. It selects the fittest individuals through a probabilistic fitness function, combining directed search with randomness to efficiently explore complex solution spaces. Various studies have used it to develop cost-based probabilistic models to determine the probable project outcomes.

For instance, Shameem et al.[44] conducted a study to address the complexity and risks associated with implementing agile methods across globally distributed software development teams. By leveraging a genetic algorithm, the authors have presented a predictive model that assesses the probability of agile project success while considering cost factors. The effectiveness of the model is validated through its ability to anticipate project outcomes by analyzing critical agile project features (factors). It used established prediction techniques, providing insights into the project features that GSD teams should prioritize to enhance the success rates of agile implementations without incurring unnecessary costs.

Alsghaier [45] presents an innovative approach to software fault prediction by combining the genetic algorithm with the support vector machine (SVM) classifier, enhanced by particle swarm optimization. This methodology aids developers in identifying faulty classes or modules early in the software development lifecycle and directs attention to areas that may require additional refactoring or testing, particularly in mission-critical systems like aircraft or medical devices. Applied to both large-scale NASA datasets and smaller Java open-source projects, the integrated approach showed improved performance in predicting software faults across diverse datasets, indicating a significant step forward in fault detection techniques and overcoming limitations found in previous studies.

Bilgaiyan et al. [46] address the challenge of effort estimation in agile software development, a significant perspective that can influence the success of a project. While agile methods are favored for their adaptability and efficiency over traditional models like waterfall and spiral, they introduce complexities in estimating costs and efforts due to their dynamic nature. Traditional estimation techniques such as analogy, expert opinion, and disaggregation lack a solid mathematical foundation. To bridge this gap, the authors propose an innovative approach using evolutionary algorithms, specifically a chaotically modified genetic algorithm (CMGA). This approach is based on mathematical morphology and utilizes a hybrid-artificial neuron model, the Dilation-Erosion perceptron (DEP), derived from complete lattice theory (CLT). The DEP-CMGA model was calibrated using data from 21 ASD projects, and its accuracy was assessed using four statistical metrics. The performance of this model was also compared with the best existing models in the literature, highlighting its potential for improving effort estimation in agile environments.

Lester et al. [47] investigate the concept of design rationale (DR), a repository of decisions and justifications made during the design process, beneficial for guiding future designs and maintaining consistency. Capturing DR is typically avoided due to its perceived cost and effort, leading to an implicit presence in various project documents. The research focuses on automating DR extraction through text mining, using machine learning to classify document sentences. It highlights the importance of feature selection in text mining and compares two evolutionary algorithms—Ant Colony Optimization (ACO) and Genetic Algorithms (GAs)—for optimizing this selection. The study assesses the effectiveness of these algorithms using the F-1 measure, aiming to determine if and how evolutionary algorithms can enhance the identification of DR from text documents.

### 7.3 Summary

Our work presents a novel contribution to the field of quantum software development by presenting a decision-making cost probability prediction model for using the traditional agile approaches to develop quantum software. Where previous studies have established the need for new frameworks and methodologies, such as the quantum agile development framework and agile methodologies tailored for quantum software, our research goes a step further by providing a detailed automated cost prediction framework for adopting traditional agile practices specifically for quantum software systems.

In the domain of using the genetic algorithm, our work stands out by applying these to develop the mentioned proposed cost prediction model, which presents cost prediction and best fitness of a set of features (causes). While other studies have successfully used genetic algorithms for predictive modeling and effort estimation in agile software development, we innovate by tailoring these algorithms to address the complexities and uncertainties of quantum software projects. This significant pivot from conventional software to quantum software represents a rise in applying established traditional computational techniques to an emerging field, potentially setting a benchmark for future research and development in quantum software engineering. Our model not only predicts project cost, best fitness of the features (causes) but also contributes to the understanding of hybrid quantum software development, thus providing a comprehensive toolset for developers and project managers in this cutting-edge domain.

# 8. Conclusion and Future Direction

Our study explored Quantum Software Engineering, a field that blends the established practices of classical software engineering with the emerging landscape of quantum computing. This integration is critical in steering architects and developers towards the creation of innovative quantum software applications, thereby advancing the field of quantum computing. The primary aim of our research was to identify key challenges and their causes in agile-based quantum software development projects and to develop a robust Agile-Quantum Software Project Success Prediction Model (AQSSPM). For this purpose, we conducted a survey to gather insights from practitioners in the field. Additionally, we jointly implemented genetic algorithms (GA) with Naive-Bayes Classifiers (NBC), and Logistic Regression (LR) to estimate the success probability of agile-quantum software projects. We also evaluated the relevance of the identified causes of the challenges in relation to their impact on the success of these projects.

Our results were quite revealing. We found that integrating GA with NBC significantly improved the probability of project success, rising from an initial 53.17% to an impressive 99.68%, and also reduced costs from 0.463% to 0.403%. In a similar vein, when GA was paired with LR, there was a notable increase in the success rate, from 55.52% to 98.99%, with a corresponding reduction in costs from 0.496% to 0.409%. Furthermore, we identified the most critical factors (causes) for the success of agile-quantum projects, which included domain-specific knowledge, effective cross-disciplinary integration, availability of resources, adaptability to technological paradigm shifts, and growing market interest. These findings are expected to provide the research and practitioner communities with robust analytical tools and strategic insights, significantly enhancing the success rates of agile-quantum projects.

Looking towards the future, we plan to test the proposed model (AQSSPM) across various quantum computing projects to evaluate its adaptability and robustness. Moreover, based on the study findings, we will enhance the model with additional predictive algorithms to improve its accuracy. Additionally, we will conduct longitudinal studies to monitor the AQSSPM's long-term performance in the fast-evolving field of quantum computing would be beneficial. Finally, we plan to develop and incorporate a structured set of guidelines specifically designed for practitioners applying the proposed model in their projects. These guidelines will include practical steps, best practices, and illustrative examples, enhancing the practical value of the model and making its accessible and actionable for real-world implementation.

## Statements and Declarations

*Acknowledgment:* The authors acknowledge that they used generative AI tools, including ChatGPT, and the writing assistant tool Grammarly, which were instrumental in addressing writing issues in this paper. It is important to note that using these tools, the authors conducted a thorough review and made substantial modifications to the content. The authors assume complete responsibility for the final content of the paper.

*Ethical Approval:* The survey data was collected based on the concent of the respondents. The sample survey questionnaire is provided at: *https://tinyurl.com/43r7uxx6*

*Competing Interests:* The authors declare that they have no known competing financial interests or personal relationships that could have appeared to influence the work reported in this paper.

*Funding:* Not Applicable.

*Availability of data and materials:* The codes and data are available under request from the authors.

*Authors' contributions:* Arif Ali Khan and Muhammad Azeem Akbar designed the study, finalize the methodology, formal analysis, and project administration. Muhammad Azeem Akbar and Valtteri Lahtinen conducted the data curation and analysis, and paper writeup done by Arif Ali Khan. Marko Paavola were mainly involved in industrial networking and polishing the final version of the article. Mahmood Niazi, Mohammed Naif Alatawi and Shoayee Dlaim Alotaibi contributed incorporating the major revision. All authors have read and agreed to the published updated version of the manuscript.

**Appendix-A:** Categories, challenges, causes and their respective best practices

| Categories [8] | Concepts (challenges(Ch)) [8] | Causes | Best practices (BP) |
|---|---|---|---|
| Knowldge and Awarness | Ch1 (Knowledge gap) <br><br> Ch2 (Team dynamics) <br><br> Ch3 (Quantum software development education) | C1 (Lack of domain specific knowledge) [BP1], [BP2], [BP3], [BP4], [BP8], [BP10] <br><br> C2 (Lack of market intrest) [BP2], [BP5], [BP6], [BP7], [BP9] <br><br> C3 (Limited acadamic reserch) [BP2], [BP5], [BP8], [BP10] <br><br> C4 (Funidng sources) [BP6], [BP7], [BP8], [BP9] | BP1- Foster collaboration and camaraderie through structured activities, aiming to enhance interpersonal relationships and team synergy. <br> BP2- Simplify complex quantum development concepts into digestible content to improve team comprehension and application skills. <br> BP3- Organize regular workshops and training sessions to raise the team's awareness and proficiency in quantum software development. <br> BP4- Implement strategies and exercises designed to solidify trust and unity among team members, crucial for tackling sophisticated projects. <br> BP5- Educate teams on the extended timelines often associated with quantum development and its potential long-term technological impact. <br> BP6- Upgrade and refine the technological infrastructure to support the unique demands of quantum software development effectively. <br> BP7- Increase awareness within the industry about the benefits and applications of quantum software, fostering broader acceptance and investment. <br> BP8- Provide support and resources for executing quantum software engineering research in academic settings to push the boundaries of the field. <br> BP9- Develop a deep understanding of how quantum software systems will interact with and affect business operations and strategies. |
| | | | BP10- Stay informed about and integrate domain-specific best practices to ensure excellence and relevance in quantum software development. |
| Sustainable scaling | Ch4 (Ethically aligned quantum software development)<br><br>Ch5 (Agile-quantum ecosystem) | C5 (Complex technical requirements) [BP13], [BP14], [BP15], [BP16]<br><br>C6 (Cross-disciplinary integration difficulties) [BP12], [BP15], [BP16], [BP17]<br><br>C7 (Rapid pace of innovation) [BP11], [BP12], [BP13], [BP14], [BP15] | BP11- Develop agile frameworks that incorporate ethical guidelines specifically designed for quantum software development, ensuring that products are aligned with societal values and norms.<br>BP12- Form agile teams comprising experts from various disciplines to facilitate the integration of diverse knowledge, which is crucial for the complex nature of quantum software development.<br>BP13- Establish innovation labs that operate within the agile methodology, dedicated to exploring and rapidly prototyping with emerging quantum technologies.<br>BP14- Implement continuous learning and development programs within agile teams to keep pace with the rapid innovation in quantum computing.<br>BP15- Create protocols that guide the effective integration of agile practices within the quantum software development process, ensuring that agility and quantum innovation complement each other.<br>BP16- Align quantum software development projects with Sustainable Development Goals to ensure that scaling up quantum computing capabilities does not come at the expense of environmental or social sustainability.<br>BP17- Introduce governance structures that apply agile principles to manage the complexity and ethical considerations of quantum software projects, allowing for flexibility and adaptability in decision-making processes. |
| Quantum-aware tools and technologies | Ch6 (Classic-quantum tailoring)<br><br>Ch7 (Continuous SE infrastructure) | C8 (Right tool for right job) [BP18], [BP20], [BP24]<br><br>C9 (Lack of industrial interest) [BP21], [BP23], [BP25]<br><br>C10 (Limited resources) [BP19], [BP21], [BP22], [BP24] | BP18- Develop and integrate agile toolkits that are tailored for quantum software development. These should facilitate the classic-quantum tailoring process and ensure that developers have the right tools for quantum-specific tasks.<br>BP19- Implement agile practices to regularly evaluate and select quantum development tools. This iterative tool assessment should be part of the sprint reviews, ensuring continuous alignment with project needs.<br>BP20- Within the agile framework, allocate resources specifically for exploring and integrating quantum technologies. This addresses the limited resources challenge by earmarking investments for quantum development.<br>BP21- Form interest groups or committees within agile teams that focus on increasing industrial interest in quantum software development through demonstrations, publications, and industry partnerships.<br>BP22- Designate specific agile sprints to focus on the transformation process from legacy software to quantum software. These sprints would deal with refactoring, integration, and the adoption of new quantum technologies. |
| | | | BP23- Conduct workshops to train agile teams on how to effectively manage the technological paradigm shift, incorporating quantum computing concepts into their existing agile practices.<br>BP24- Develop CI/CD pipelines that are capable of supporting both classical and quantum software development, ensuring that agile practices can sustain a continuous SE infrastructure.<br>BP25- Encourage agile teams to engage in technological foresight to anticipate and prepare for future shifts in quantum technology, integrating this into their sprint planning and backlog refinement. |
| Standards & Specifications | Ch8 (Processes standardization)<br><br>Ch9 (Optimum documentation) | C11 (Varying Interpretations of Agile Methodologies) [BP29], [BP32]<br><br>C12 (Rapid Evolution of Quantum Technologies) [BP27], [BP28], [BP31], [BP32], [BP34]<br><br>C13 (Complexity of Quantum Computing Concepts) [BP29], [BP31], [BP33]<br><br>C14 (Necessity of Integration with Existing Standards) [BP26], [BP27], [BP28], [BP30], [BP31] | BP26- Develop maturity models specific to agile-quantum development to assess and guide teams on their journey towards optimized agile practices in quantum software creation.<br>BP27- Institute standard assessment protocols that evaluate the effectiveness of agile practices and documentation in quantum development, ensuring they meet the high-quality standards required in this field.<br>BP28- Create a comprehensive integration framework that lays out best practices for merging agile methodologies with quantum software development, including checklists and templates for standardized processes.<br>BP29- Implement quality standards for documentation that are specifically designed for the quantum software domain, guiding teams on creating clear, concise, and useful documentation artifacts.<br>BP30- Integrate sprints that are dedicated to ensuring the project's compliance with established frameworks and standards, allowing teams to correct course and align with best practices throughout the development cycle.<br>BP31- Encourage the use of SDKs that come with built-in standards and documentation guidelines, making it easier for agile teams to adhere to best practices without extensive overhead.<br>BP32- Compile playbooks that detail the agile-quantum processes, offering standardized procedures and documentation templates that can be customized for various quantum development scenarios.<br>BP33- Conduct focused workshops to train agile teams on creating tailored documentation that meets the specific needs of quantum software development while adhering to agile principles. These workshops would emphasize the creation of just-enough and just-in-time documentation, ensuring relevance and agility.<br>BP34- Dedicate specific sprints to the development and refinement of documentation, ensuring that it keeps pace with the rapid development cycles typical in quantum software projects. These sprints would focus on updating documentation to reflect new learnings, architectural decisions, and code changes, maintaining an optimum level of documentation throughout the project lifecycle. |

| Common Causes Accross All the Categoreis | C15 (Technological Paradigm Shift) [BP35], [BP38], [BP40], [BP41]<br><br>C16 (Transforming process from legacy to quantum software) [BP35], [BP36], [BP38]<br><br>C17 (Gap between research and practice) [BP35], [BP37], [BP40]<br><br>C18 (Emerging working culture) [BP41], [BP42], [BP43]<br><br>C19 (Interdisciplinary Collaboration Barriers) [BP39], [BP40], [BP42] | BP35- As quantum computing is a technological paradigm shift, it is essential to invest in continuous learning opportunities for team members to keep up with the latest quantum computing developments and practices. Regular training sessions, workshops, and seminars can help bridge the gap between research and practice.<br>BP36- When moving from legacy to quantum software, employ an incremental approach that allows for iterative testing, validation, and integration of quantum components.<br>BP37- Form alliances with academic and research institutions to stay abreast of the latest quantum computing research and to help translate those findings into practical applications.<br>BP38- Implement an agile methodology that is flexible enough to accommodate the rapid changes inherent in quantum technology and allows for adapting workflows to integrate new findings and techniques.<br>BP39- Create cross-functional teams with diverse skill sets, including quantum physicists, software developers, and domain experts, to foster interdisciplinary collaboration and innovation.<br>BP40- Develop internal knowledge-sharing platforms to disseminate quantum computing knowledge and practices within the organization, bridging the gap between different disciplines.<br>BP41- Encourage a culture that embraces experimentation and tolerates failure, recognizing that the quantum field is exploratory and that failure can provide valuable insights.<br>BP42- Utilize modern communication and collaboration tools to support the emerging working culture, which may be more distributed and remote, particularly in the interdisciplinary field of quantum computing.<br>BP43- Employ agile project management techniques specifically tailored for quantum software development projects, accounting for their high uncertainty, rapid evolution, and the need for close collaboration across different fields. |